# Unraveling Bulk and Grain Boundary Electrical Properties in $La_{0.8}Sr_{0.2}Mn_{1-y}O_{3\pm\delta}$ Thin Films

Francesco Chiabrera[1], Iñigo Garbayo[1], Dolors Pla[2], Mónica Burriel[2], Fabrice Wilhelm[3], Andrei Rogalev[3], Marc Núñez[1], Alex Morata[1], Albert Tarancón[1,4,*]

[1]Department of Advanced Materials for Energy, Catalonia Institute for Energy Research (IREC), Jardí de les Dones de Negre 1, Planta 2, 08930 Sant Adrià de Besòs (Barcelona), Spain.

[2]Univ. Grenoble Alpes, CNRS, Grenoble INP,[†] LMGP, F-38000 Grenoble, France

[3]European Synchrotron Radiation Facility (ESRF), F-38054 Grenoble, France.

[4]ICREA, Passeig Lluís Companys 23, 08010 Barcelona, Spain

* Corresponding author´s e-mail: atarancon@irec.cat;

**Abstract**

Grain boundaries in Sr-doped $LaMnO_{3\pm\delta}$ thin films have been shown to strongly influence the electronic and oxygen mass transport properties, being able to profoundly modify the nature of the material. The unique behaviour of the grain boundaries can be correlated with substantial modifications of the cation concentration at the interfaces, which can be tuned by changing the overall cationic ratio in the films. In this work, we study the electronic properties of $La_{0.8}Sr_{0.2}Mn_{1-y}O_{3\pm\delta}$ thin films with variable Mn content. The influence of the cationic composition on the grain boundary and grain bulk electronic properties is elucidated by studying the manganese valence state evolution using spectroscopy techniques and by confronting the electronic properties of epitaxial and polycrystalline films. Substantial differences in the electronic conduction mechanism are found in the presence of grain boundaries and depending on the manganese content. Moreover, the unique defect chemistry of the nanomaterial is elucidated by measuring the electrical resistance of the thin films as a function of oxygen partial pressure, disclosing the importance of the cationic local non-stoichiometry on the thin films behavior.

---

[†] Institute of Engineering Univ. Grenoble Alpes

Engineering the transport properties of complex oxides is considered one of the major challenges for launching the next generation of highly performing solid state devices. In this direction, mastering the implementation of interface-dominated materials represents a powerful engineering tool with implications in many different fields. For instance, grain boundaries (GBs) and dislocations have been demonstrated to substantially increase oxygen diffusion in $La_{0.8}Sr_{0.2}MnO_{3\pm\delta}$ (LSM) thin films, improving their performance as cathodes in solid oxide fuel cells.[1–4] Opposite, GBs are also responsible of a strong reduction of magnetic order and electronic structure in manganites, which can be used in spintronics and resistive switching devices.[5,6] In this sense, GBs are generally considered as an insulating paramagnetic region, whose presence can give rise to a collection of unique properties, such as a large magnetoresistance below Curie temperature[7,8] or the appearance of different low temperature conduction mechanisms in polycrystalline thin films.[9,10]

We recently showed that the GBs in LSM thin films are dominated by an unusual local modification of chemical composition, which involves both the oxygen and the cations.[11] Moreover, we demonstrated that this local non-stoichiometry is modifiable by simply changing the overall cationic ratio Mn/(La+Sr) in the thin films. This way, for highly Mn deficient thin films the GBs ($GB_{def}$) are characterized by an additional depletion of Mn and O, while La and Sr increase. When the overall Mn concentration rises ($GB_{rich}$), the interfaces completely change the cationic distribution, turning to La (and O) deficient regions with progressive increase of Mn. Notably, we confirmed that the variations observed in the GB composition are the cause of strong changes in the electrical and electrochemical behaviour, thus offering a powerful tool for tuning the functional properties of thin films.

In this work, we delve into the electronic conduction mechanism of $La_{0.8}Sr_{0.2}Mn_{1-y}O_{3\pm\delta}$ ($LSM_y$) thin films by studying, first, the impact of the different Mn/(La+Sr) ratios on the bulk behaviour by spectroscopy techniques and, second, the GB properties in polycrystalline samples by confronting them with epitaxial $LSM_y$ with variable Mn content. The first analysis on the bulk defect chemistry is particularly important since large Mn deficiencies are explored in this study. In this sense, it is well-known that a change in the B/A ratio creates negative cation vacancies in LSM ($V'''_{Mn}$ or $V'''_{La}$) that need to be compensated by positive defects ($V_O^{\cdot\cdot}$ and/or $Mn^{4+}$), strongly affecting the electronic conduction properties. Using this first analysis for defining the bulk references, we aim to identify then the distinct conduction mechanism occurring at the GBs, stressing the importance of these interfaces in the overall functional properties of thin films. Finally, we discuss the impact of the specific local non-stoichiometry at the GBs on the electronic mechanism, showing the central role of Mn in defining the interface properties.

$LSM_y$ thin films with variable Mn content were deposited by combinatorial Pulsed Laser Deposition (c-PLD) on 5mm x 10mm Sapphire (0001) substrates and on 100 nm YSZ / Si 4" wafers in a large-area system from PVD Products (PLD-5000), equipped with a KrF-248 nm excimer laser from Lambda Physics (COMPex PRO 205). C-PLD is a technique based on the alternate deposition of two different parent compounds (in our case $La_{0.8}Sr_{0.2}MnO_3$ and $Mn_3O_4$ commercial targets) at two opposite sides of a large area substrate.[12–15] Due to the typical Gaussian shape of the PLD plasma plumes, a continuous gradient of concentration is obtained between the two plumes centres. For the $LSM_y$ combinatorial deposition, the sapphire samples were pasted to a 4" Si wafer forming a line between the LSM and $Mn_3O_4$ plume centers, allowing a gradual increase of B/A=Mn/(La+Sr) ratio. The LSMy / YSZ / Si sample was instead deposited at wafer level and then cut into 1x0.5 $cm^2$ rectangular specimens for the electrical and structural characterization.[11] Deposition parameters were adapted in order to obtain fully dense films with compositions from Mn/(La+Sr) ratio of B/A = 0.85 ± 0.02 to B/A > 1.2 ± 0.03, which was confirmed by Wavelength Dispersive Spectroscopy and Energy Dispersive X-ray Analysis, and thicknesses between 70 and 100 nm (see details elsewhere[11]). The B/A compositional error in the structural and electrical characterization was calculated considering the variation of B/A ratio occurring in the portion of combinatorial sample analysed.[11] In order to equilibrate the oxygen stoichiometry of the samples, all the $LSM_y$ thin films were annealed for 6 hours in air at 923 K after deposition. Structural and surface characterization of the thin films was performed by X-ray Diffraction (XRD) (Bruker D8 Advance diffractometer system), Transmission Electron Microscopy (TEM) (FEI TITAN Low Base operated at 300 kV) and by non-contact Atomic Force Microscopy (AFM, XE 100 Park System Corp.). **Figure 1(a)** shows the Θ-2Θ XRD of the $LSM_y$ samples revealing a single phase deposition independent of the Mn content. These measurements were performed applying an omega offset to reduce the contribution from the single crystalline substrates. A slight $(110)_{pc}$ preferential orientation is detected in the most deficient Mn thin films, which reduces increasing the B/A ratio and disappear for the nearly stoichiometric layer. The Mn rich samples show instead a small strengthening of $(100)_{pc}$ orientation. **Figure 1(b)** shows the evolution of the lattice parameters as a function of the B/A ratio. A significant linear expansion of the unit cell is observed in the Mn deficient part, while a small shrinkage is detected increasing the Mn content (B/A > 1), as previously reported.[11] The $LSM_y$ films deposited on Sapphire and YSZ/Si show comparable lattice parameters, implying that the substrate is not affecting the stress state in the polycrystalline layers. At the same time, AFM surface and TEM cross section images of a typical $LSM_y$ sample with B/A = 0.92 ± 0.02, **Figure 1(c)** and **(d)** respectively, prove the characteristic columnar structure and nanocrystalline grains obtained. This columnar structure, without segregation of secondary phases, was observed in all the samples independently of the overall Mn content, with a small variation of the grain size for different B/A ratios.[11]

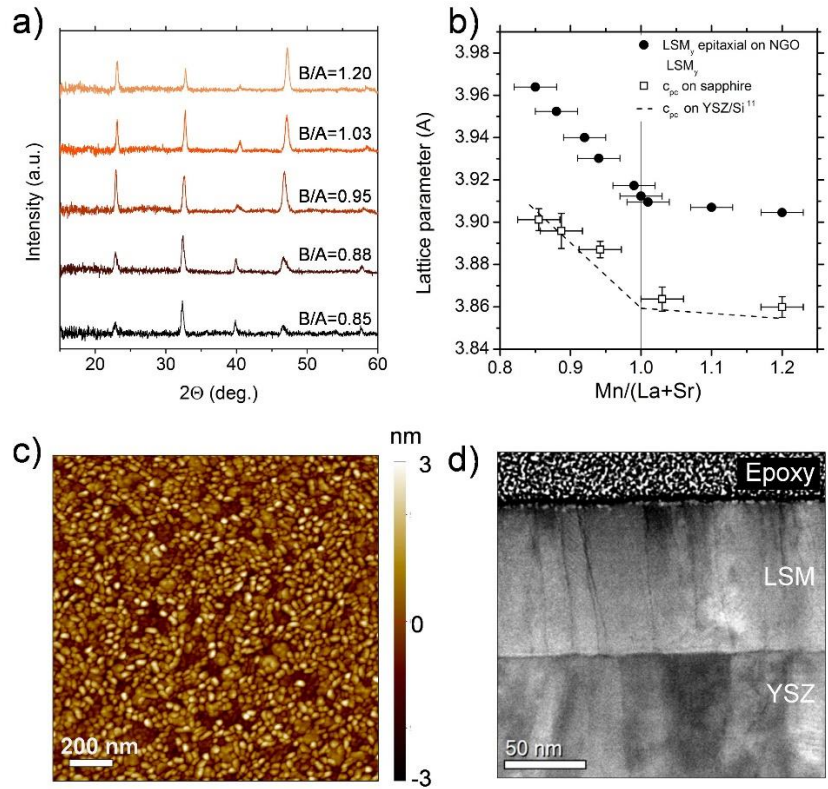

**Figure 1.** (a) XRD of the polycrystalline $LSM_y$ thin film grown on the sapphire (0001). (b) Evolution of the pseudo-cubic lattice parameter as a function of the B/A ratio derived from the XRD analysis for different substrates (c) AFM image of the surface and (d) cross-section dark field TEM image of a characteristic $LSM_y$ sample deposited on a 100 nm YSZ / Si substrate.

X-ray Absorption Near-Edge Structure (XANES) and ellipsometry were carried out on the LSM$_y$ film series in order to investigate the effect of varying the B/A ratio on the overall manganese valence. It is important to note that these two techniques are not able to separate grain and grain boundary contributions, therefore the results described next can be unambiguously ascribed to the bulk LSM$_y$, since it represents approximately the 90 vol.% of the polycrystalline films. XANES spectra at the Mn K-edge of LSM$_y$ thin films were collected at the ESRF ID12 Synchrotron beamline (Grenoble, France) under vacuum and at T = 298 K with a nearly constant beam current of 200 mA (see additional details in **Section S.1.** of **supplementary information**). **Figure 2(a)** shows the Mn K-edge XANES spectra of the polycrystalline LSM$_y$ thin films for different B/A ratios. Here we focus on the absorption edge region, particularly on the white line positioned at *ca.* 6554 eV. Its position for each spectrum was extracted by calculating the inflection point energy, given by the maximum of the first derivative (see **Figure S1** of the **supplementary information**). This feature, which involves a transition from Mn 1s to Mn 4p states, has been demonstrated to display an energy shift depending on the Mn valence in many different compounds, including perovskites.[16,17] In particular, the K-edge energy increases almost linearly with the Mn valence in LSM, shifting approximately 1 eV per 0.27 valence change.[18,19] In our samples, no significant changes were measured in the Mn deficient films and only a very small increase of ~ 0.1 eV (corresponding to a relative increase of 3% in $[Mn^{4+}]$[18–20]) was observed in the La-rich compositions (**Figure 2(b)**). Complementary, ellipsometry measurements supported the XANES results by confirming a uniform Mn valence within the Mn deficient LSM$_y$ films (see detailed analysis in **Section S.2.** of the **supplementary information**). Interestingly, the optical conductivity measured shows a high sensitivity to the Mn content, offering a powerful tool to track compositional changes by studying the optical properties of the material.

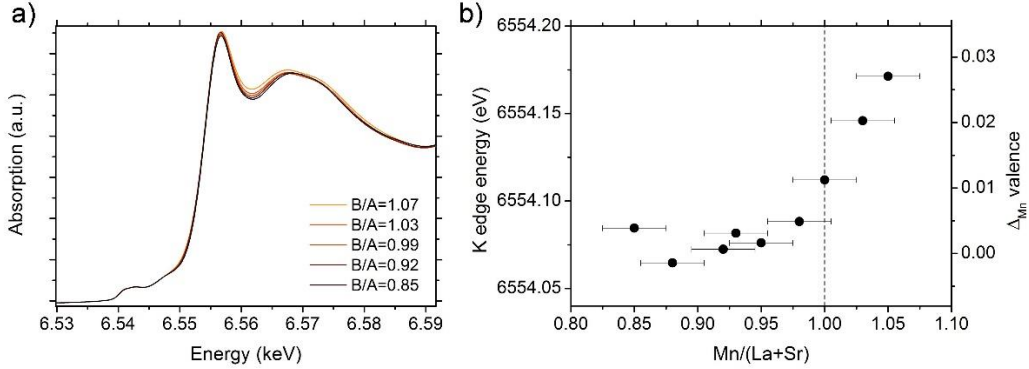

**Figure 2.** (a) Normalized XANES spectra at the Mn K-edge for various B/A ratios. (b) Variation of the calculated inflection points energy with the cationic composition in LSM$_y$ films. The corresponding relative change of Mn valence is also shown in the right axis.[18,19]

According to the XANES and ellipsometry results, the bulk electro-neutrality in the Mn-deficient LSM$_y$ films cannot be attributed to a change in the Mn valence state. The most probable compensation mechanism would be the formation of positive defects, such as oxygen vacancies, usually present in a very low concentration in stoichiometric LSM.[21,22] However, the total charge compensation by oxygen vacancies (with equilibrium $[V_O^{\cdot\cdot}] = 3/2 \cdot [V_{Mn}''']$) would result in a great oxygen deficiency and a much higher increase of bulk oxygen diffusion than the one observed in our previous works.[1,3] Alternatively, we propose a different mechanism in which the cation vacancies are mainly compensated by the formation of neutral anti-sites defects ($La_{Mn}^x$). Simple calculations after the work of Nakamura *et al.*[23] (included in **Section S.3.** of **supplementary information**) show that this compensation mechanism radically reduces the need of oxygen vacancies to balance the charge; *e.g.* for the nominal composition B/A = 0.90, a concentration of $La_{Mn}^x = 0.04$ lowers the effective concentration of $V_{Mn}'''$ from 0.10 to nearly 0.02 per unit-formula, implying that significantly less negative charge needs to be compensated by positive defects (i.e. oxygen vacancies). Although this mechanism is not commonly observed in bulk perovskites, similar anti-site defects have been recently reported by Qiao *et al.* in Cr deficient epitaxial LaCrO$_3$ films.[24] Moreover, recent Density Functional Theory (DFT) simulations carried out for LSM[11] showed that the formation energy of $La_{Mn}^x$ strongly decreases in the presence of oxygen vacancies, i.e. the formation of the anti-site defects is favourable under highly reducing PLD growing conditions. This compensation mechanism is also in good agreement with the expansion of the out-of-plane lattice parameter measured in our films (see **Figure 1(b)**), since the La ionic radius is significantly bigger than the Mn one. As a counterpart, for moderately Mn rich samples (B/A > 1), the presence of La vacancies ($V_{La}'''$) can be mainly compensated by the formation of the likely occurring $Mn_{La}^x$ anti-site defects accompanied by a small increase of $Mn^{4+}$, which can explain the slight shrinkage of the unit cell observed (see **Figure 1(b)**).

In order to clearly identify the contribution of GBs on the electrical properties of polycrystalline LSM$_y$ films, we compared two sets of equivalent polycrystalline (on sapphire (0001) and on YSZ/Si substrates) and epitaxial (on NdGaO$_3$ (110), NGO) samples with B/A ratios from 0.85 to 1.2. The epitaxial films showed a single phase, (001)$_{pc}$ oriented growth, with a smooth surface of rms < 0.4 nm. The evolution of out of plane lattice parameter with the B/A ratio follows the same trend of the polycrystalline layers, see **Figure 1(b)** (the complete structural and compositional characterization of the epitaxial films can be found in **Section S.4.** of the **supplementary information**). All the samples were electrically characterized in air in a 4-point in-plane configuration by using silver electrodes in a temperature-controlled Linkam probe station, between 83 K and 873 K (in conformity with our previous works, see refs.[3,11] and **Section S.5.** of the **supplementary information**).

First, the in-plane electronic conductivity above room temperature of polycrystalline and epitaxial films with different Mn content have been represented using an Arrhenius plot in **Figure 3(a)**. In general, the polycrystalline thin films present lower conductivity with respect to the epitaxial ones of similar composition, while an increase in conductivity is observed by increasing the Mn content, for both types of films. Independently of the composition or nature of the films, these show a perfect agreement with a small polaron hopping conduction for the whole range of temperatures,[10,22,25–27] also supported by the optical measurements (see **Figure S2** in **supplementary information**). According to this model, the polarons can move to the Nearest Neighbour Hopping (NNH) site with the assistance of phonons, leading to the following expression of conductivity in the case of adiabatic limit:[28]

$$\sigma(T) = \frac{\sigma_0}{T} \cdot \exp\left(\frac{E_a}{k_b T}\right) \quad (1)$$

Where $E_a$ refers to the activation energy of the hopping mechanism, $k_b$ is the Boltzmann constant and $\sigma_0$ is the pre-exponential factor.

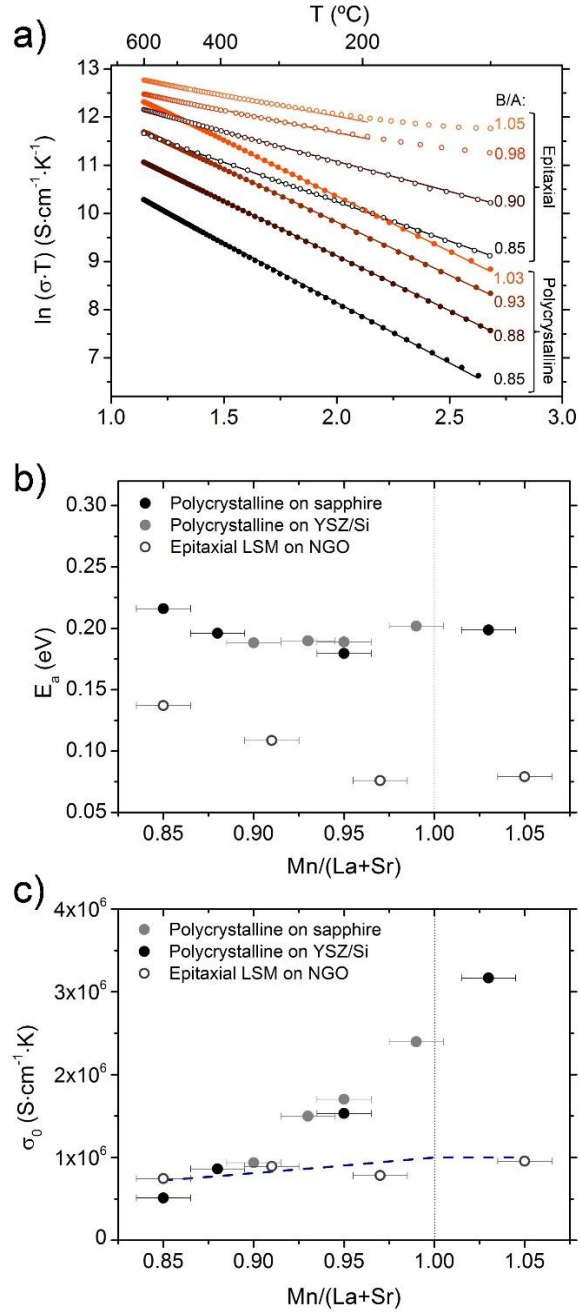

**Figure 3.** (a) In-plane electronic conductvity as a funcion of temperature for the epitaxial and polycrystalline LSM$_y$ thin films, plotted according to the adiabatic small polaron model. (b) Energy of activation ($E_a$) and (c) pre-exponential factor ($\sigma_0$) derived by the small polaron fitting. The polycrystalline samples measured on two different substrates give rise to similar results. The blue dashed line in (c) indicates the theoretical trend expected from the decrease of Mn hopping sites and no change in Mn valence ($\sigma_0 \propto [Mn]^2$).

By fitting the conductivity values with Equation 1, $E_a$ and $\sigma_0$ have been calculated and represented in **Figure 3(b)** and **(c)**, respectively. First, the activation energy of the epitaxial samples is lower than that of the polycrystalline ones. Moreover, the epitaxial samples present a progressive decrease of the activation energy while increasing the B/A ratio, whereas in the polycrystalline ones $E_a$ is kept almost constant. This can be clearly associated to the presence of structural and chemical defects in both types of GBs ($GB_{def}$ and $GB_{rich}$) that cause the localization and trapping of charge carriers, breaking the conduction paths and increasing the polaron energy barrier. Noteworthy, the decrease of $E_a$ in the epitaxial samples is consistent with the decrease observed in the out-of-plane lattice parameter (see **Figure 1(b)**), which represents a reduction of the tetragonal distortion of the cell and diminishes the polaron barrier.[29–31] Also, this matches with the shift towards lower energies for both, the polaron and Jahn-Teller distortion phenomena, which govern the optical transitions observed by ellipsometry when increasing the Mn content in the films (see **Figure S2** in **supplementary information**).[32]

**Figure 3(c)** shows the evolution of the pre-exponential factor with the B/A ratio. In the epitaxial films, a small increase could be measured with increasing the Mn content in the films. The pre-exponential factor is defined in the case of considering on-site coulombic repulsions as:[31,33]

$$\sigma_0 = \frac{e^2 \nu}{a k_b} \cdot [Mn^{4+}] \cdot [Mn^{3+}] \tag{2}$$

Where e, $\nu$, $a$ are the electron charge, the polaron attempt frequency and the site to site hopping distance, respectively. In the grain bulk, since negligible manganese valence changes were measured, the pre-exponential directly increases with the Mn content, $\sigma_0 \propto [Mn^{4+}] \cdot [Mn^{3+}] \propto [Mn]^2$. Thus, a decrease of the Mn concentration generates fewer sites for the polarons to hop, resulting in lower values of conductivity. As a matter of fact, the pre-exponential factor of the epitaxial LSM$_y$ films perfectly follows this theoretical trend (blue dashed line in **Figure 3(c)**), reinforcing the proposed defect model for the bulk material. Even more interestingly, polycrystalline films present a surprising increment of $\sigma_0$ as a function of Mn (from $GB_{def}$ to $GB_{rich}$) indicating the great impact of the cation non-stoichiometry at the GB level. While in the presence of $GB_{def}$ the pre-exponential factor is slightly smaller than the bulk (straightforwardly related to the extra Mn depletion at the GB), in the films dominated by $GB_{rich}$ the increase is much higher suggesting the creation of additional pathways for electronic conduction probably enabled by the generation of anti-site $Mn_{La}^x$ at the interface level.[34] The complementary $\sigma_0$ and $E_a$ measured for the polycrystalline thin films deposited on YSZ/Si and on sapphire suggest that the substrate does not influence the GB composition and, therefore, the electrical properties. All in all, a remarkable impact of the GBs on the high temperature electrical conduction of polycrystalline samples is clearly proved.

Complementary to **Figure 3(a)**, **Figure 4(a)** shows a representation of the conductivity of specific compositions of epitaxial and polycrystalline thin films according to the NNH conduction model, for the whole range of temperatures measured, namely, from 83 K to 873 K. According to this figure, nearly stoichiometric (B/A = 1.03) polycrystalline layers, as well as epitaxial thin films (both Mn-deficient, B/A = 0.85 and slightly Mn-rich, B/A=1.05), present the typical metal-insulator (M-I) transition observed for LSM.[35] The less marked metallic behaviour of the epitaxial Mn deficient epitaxial sample can be explained by the presence of scattered Mn vacancies, which are known to hinder the long range electronic order in manganites.[36–38] Opposed to that, a suppression of the M-I transition, maintaining the insulating behaviour, is observed for the polycrystalline Mn deficient LSM$_y$, see further details in reference [11]. Moreover, its conductivity deviates from the linear behaviour expected for the NNH model, approaching the typical performance of a Variable Range Hopping (VRH) conduction mechanism, **Figure 4(b)**, in which the probability of finding an available site with a proper energy level decrease with temperature, determining a temperature-dependent hopping length.[39] VRH theory was first developed for disordered materials and follows the general expression:

$$\sigma(T) = \sigma_p \cdot \exp\left(-\frac{T_0}{T}\right)^p \tag{3}$$

With $T_0$ being the characteristic temperature and p the exponential term, that equals to ¼ in the case of a Mott regime[39] and ½ for a Shklovskii–Efros (SE-VRH) mechanism.[40] For Mn-deficient polycrystalline LSM films, we found a good consistency with the SE-VRH model (see **Figure 4(b)** and more details in **Section S.6.** of the **supplementary information**). The appearance of VRH conduction in the polycrystalline films with $GB_{def}$ is not intrinsic but due to the localized defects in the GBs, since the epitaxial film with the same composition shows a better agreement with the NNH adiabatic model in the whole temperature range above the occurrence of the M-I transition. Indeed, the onset of this mechanism was also previously reported in polycrystalline manganite thin films,[9,10] as well as in disordered bulk manganites[41,42] and ultrathin epitaxial thin films.[43] From the mechanistic point of view, the Mn deficiency and the presence of oxygen vacancies at the interface in $GB_{def}$ of polycrystalline $LSM_y$ films naturally hinder the insulator metallic behaviour giving rise to the VRH mechanism, in which polarons need to jump towards distant sites across the GB. The appearance of the SE-VRH could indicate that the charged defects in the GBs promote polarons trapping or that the density of states at the fermi level is not constant (see **Section S.6.** in the **supplementary information** for additional details on the ES-VRH model).[41,42,44] The recovery of the M-I transition for $GB_{rich}$ films indicates that the depletion of La is not playing an active role in the electronic conduction and that the oxygen defect equilibrium of these GBs is similar to the one of the bulk.

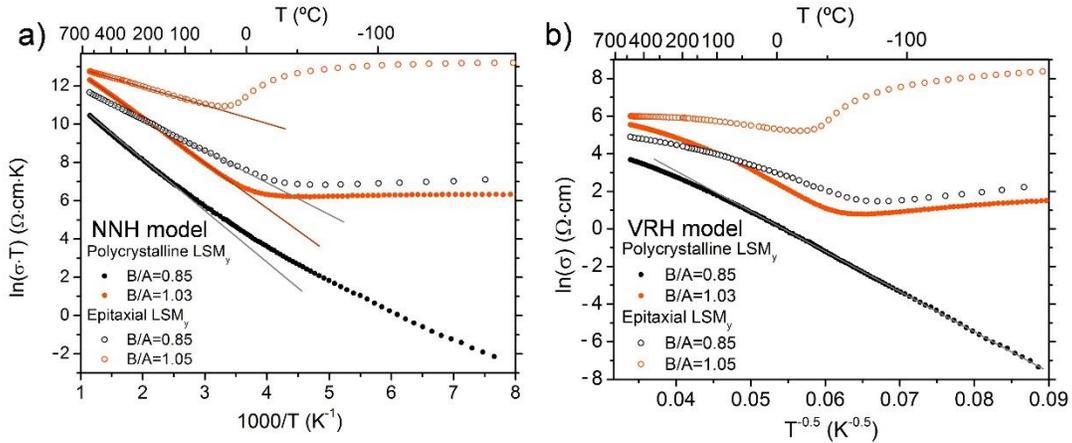

**Figure 4.** In plane electronic conductvity plotted according to the adiabatic small polaron model (a) and to the SE-VRH model (c) for a wide set of temperatures. In the polycrystalline sample with B/A = 0.85 the first well describe the conductivity for T > 473 K, while for T < 473 K the second shows a better fitting.

At this stage, it is clear that GBs in $LSM_y$ films are in the origin of significant changes in the electrical properties of the films ($E_a$, $\sigma_0$ and suppression of M-I transition), and that a distinct conduction mechanism takes place, presumably linked to previously reported occurring local stoichiometric changes.[11] These, most likely modify the defect chemistry of the bulk material (Mn deficient or not). In order to get deeper insights into the defect chemistry of $LSM_y$ grain boundaries, we analysed the high temperature oxygen partial pressure (pO$_2$) dependence of the in-plane electronic conductivity. The experiments were performed in a ProboStat test station (NorECs). The pO$_2$ in the station was controlled with a ZIROX oxygen pump and monitored with an oxygen lambda-sensor placed next to the sample. The temperature was set to 923 K for ensuring a fast equilibration with the atmosphere while preventing grain growth in the polycrystalline films.

**Figure 5(a)** shows the $pO_2$ dependence of the electrical conductivity for polycrystalline thin films with different compositions. The conductivities were normalized to the value measured at 50 ppm of oxygen to better visualize the relative changes originated by the $pO_2$, since the Mn deficient samples show a much lower conductivity respect to the stoichiometric ones (see **Figure S7** in **supplementary information**). Results obtained for an epitaxial film with nominal composition B/A = 0.85 are also included for comparison. Interestingly, the Mn-deficient epitaxial sample displays the typical p-type behaviour expected for the bulk defect chemistry while for the polycrystalline samples p-type and n-type like dependences are observed for B/A > 0.9 ($GB_{rich}$) and B/A < 0.9 ($GB_{def}$), respectively. Oxidation-reduction studies of Mn-deficient polycrystalline samples shown in **Figure 5(b)** prove the reversibility of the unusual n-type like response, discarding any type of stabilization issues and confirming that the phenomenon is directly linked to the high concentration of GBs present in the films. Moreover, complementary measurements on thicker samples with lower GB density (double grain size) corroborate the dominant role of the interfaces in the conductivity (**Figure S8** in **supplementary information**), since two contributions could be clearly differentiated when changing $pO_2$, namely, a fast one with the unusual n-type like behaviour attributable to the $GB_{def}$ and a slower one showing the expected p-type conduction of the grain bulk.

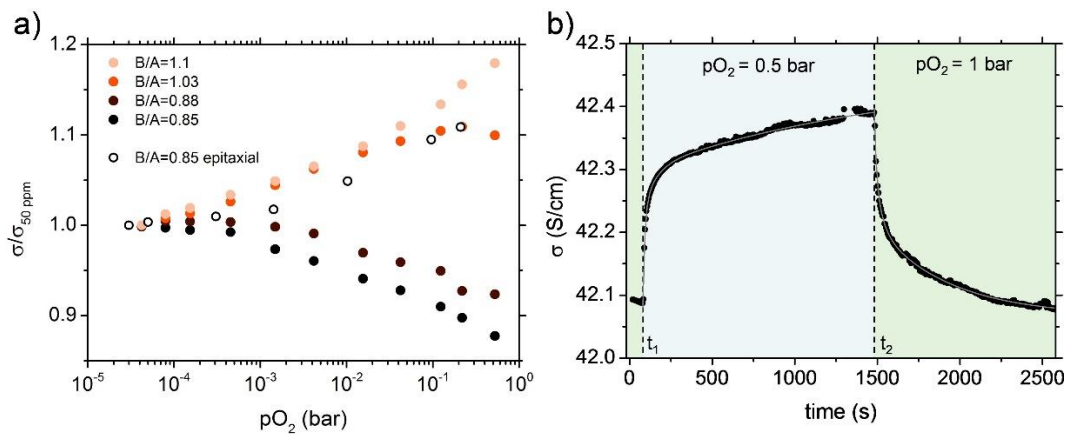

**Figure 5.** (a) Relative in-plane electronic conductivity of the polycrystalline LSM$_y$ films as a function of the $pO_2$ (vs. the low $pO_2$ (50 ppm) conductivity) measured at 923 K. The epitaxial film with B/A = 0.85 is also reported for comparison. (b) Time response of the conductivity for reducion and oxidation steps at 923 K in the polycrystalline thin films with B/A = 0.85. The lines indicate the times at which the oxygen content in the atmosphere were changed ($t_1$ and $t_2$).

As previously mentioned, the conductivity of $GB_{rich}$ films follows the bulk behavior of LSM suggesting a similar defect chemistry, *i.e.* incorporating oxygen by the creation of $V_{Mn}'''$ and/or $V_{La}'''$ combined with an increase of $[Mn^{4+}]$ that compensates the negative charges, resulting in the slight p-type oxygen dependence of the electronic conductivity.[45] Contrary, the origin of the unusual pO$_2$ dependence of $GB_{def}$ films is somehow more controversial. First, it could be due to the appearance of $Mn^{2+}$ in the GB core, which would give rise to a n-type conduction even in oxidizing conditions.[21] However, our previous studies contradicts this hypothesis since we did not observe any local change of oxidation state in $GB_{def}$.[11] Another possibility would be the creation of a negative space charge layer caused by the positive oxygen vacancies in the GB core, which could originate an enrichment of negative elements ($Mn^{2+}$) in the GB surroundings.[46] Additionally, the presence of negative cation defects, which are also actively involved in the oxidation mechanism, could also play a role in defining the space charge properties.[47] Nevertheless, due to the high carrier concentration of the bulk and the small band-gap of LSM, the expected width of the space charge layer is extremely small, apparently ruling out this possibility.[7,48] Finally, a mechanism based on trapping positive charge carriers created upon oxygen oxidation may explain the pO$_2$ trend observed. Indeed, an increase of $Mn^{4+}$ necessarily implies a decrease of $Mn^{3+}$ and, according to equation 2, this could originate a decrease of the total conductivity (if $Mn^{4+}$ is trapped). In any case, the appearance of this unusual pO$_2$ behaviour is undoubtedly related to the characteristic depletion of Mn in $GB_{def}$ and could have direct implications especially on the electrochemical performance of the layers and their stability in redox conditions. In this sense, we previously observed an outstanding stability of Mn deficient polycrystalline LSM layers when employed in symmetrical cells resulting in an anomalous weak dependence with pO$_2$ of their oxygen diffusivity,[49] contrary to what is expected from the defect chemistry models.[21]

In summary, we were able to fabricate high quality epitaxial and dense polycrystalline thin films of La$_{0.8}$Sr$_{0.2}$Mn$_{1-y}$O$_{3\pm\delta}$ with B/A ratio ranging from 0.85 (Mn deficient) to 1.2 (La deficient films). Conductivity measurements of epitaxial and polycrystalline LSM$_y$ thin films allowed us to elucidate the unique features of the GBs. From our study, it is clear that the electronic conductivity of the polycrystalline samples is dominated by the GBs, which largely impact the activation energy, the pre-exponential factor and even the conduction mechanism at low temperatures. The origin of these effects is ascribed to the local non-stoichiometry at the GB level reported by our group in a previous work. Moreover, an unusual $pO_2$ dependency of the conductivity was observed as a function of the Mn content, indicating that different defect chemistries are required to fully describe the GB oxygen incorporation in $LSM_y$ films with different B/A ratio. All in all, the results of this work, combined to the recently discovered fast ion conduction along LSM GBs, anticipate a radical change in the understanding and implementation of complex oxides in a collection of thin film-based solid state devices, especially for energy and information applications.

## Acknowledgements

The research was supported by the Generalitat de Catalunya-AGAUR (2017 SGR 1421). This project has received funding from the European research Council (ERC) under the European Union's Horizon 2020 research and innovation programme (ULTRASOFC, Grant Agreement number: 681146) and under the Marie Skłodowska-Curie Grant Agreement 746648 — PerovSiC (for D.P). The XANES experiments were performed on beamline ID12 at the European Synchrotron Radiation Facility (ESRF), Grenoble, France. We are grateful to Fabrice Wilhelm at the ESRF for providing assistance in using beamline ID12.

## Competing interests

The authors declare no competing interests.

## Supplementary Information

See supplementary information for Supplementary Figures S1–S7, Supplementary References 1–26.